\documentclass[journal, letter, onecolumn]{IEEEtran}
\usepackage{setspace}

\usepackage{graphicx}
\usepackage{epsfig}
\usepackage{latexsym}
\usepackage{amsfonts}
\usepackage{here}
\usepackage{rawfonts}
\usepackage[latin1]{inputenc}
\usepackage[T1]{fontenc}
\usepackage{calc}
\usepackage{url}
\usepackage{enumerate}
\usepackage{color}
\usepackage[tbtags]{amsmath}
\usepackage{amssymb}
\usepackage{upref}
\usepackage{epic,eepic}
\usepackage{times}
\usepackage{dsfont}
\usepackage{comment}
\usepackage{cite}

\usepackage{graphicx}
\usepackage[font={small}]{caption}
\usepackage[font={small}]{subcaption}

\usepackage{stfloats}
\usepackage{mathtools}
\usepackage{amsmath}

\usepackage{algorithm}
\usepackage{algpseudocode}

\newtheorem{theorem}{\bf{Theorem}}
\newtheorem{definition}{\bf{Definition}}

\ifCLASSINFOpdf
\else
\fi
\begin{document}

\title{Mat{\'e}rn Cluster Process with Holes at the Cluster Centers}

\author{{Seyed Mohammad Azimi-Abarghouyi} and Harpreet S. Dhillon

	\thanks{S. M. Azimi-Abarghouyi is with the School of Electrical Engineering and Computer Science, KTH Royal Institute of Technology, Stockholm, Sweden (Email: seyaa@kth.se). H. S. Dhillon is with Wireless@VT, Bradley Department of Electrical and Computer Engineering, Virginia Tech, Blacksburg, VA USA (Email: hdhillon@vt.edu). }
}


\maketitle

\begin{abstract}
Inspired by recent applications of point processes to biological nanonetworks, this paper presents a novel variant of a Mat{\'e}rn cluster process (MCP) in which the points located within a certain distance from the cluster centers are removed. We term this new process the {\em MCP with holes at the cluster center} (MCP-H, in short). Focusing on the three-dimensional (3D) space, we first characterize the conditional distribution of the distance between an arbitrary point of a given cluster to the origin, conditioned on the location of that cluster, for both MCP and MCP-H. These distributions are shown to admit remarkably simple closed forms in the 3D space, which is not even possible in the simpler two-dimensional (2D) case. Using these distributions, the contact distance distribution and the probability generating functional (PGFL) are characterized for both MCP and MCP-H. 

\end{abstract}

\section{Introduction}
Clustered point patterns appear in many diverse areas of science and engineering, such as geodesy, ecology, biology, and wireless networks. Owing to their generality and tractability, Poisson cluster processes (PCPs) are often the first choice for modeling such point patterns. As a representative application, PCPs \cite[Sec. 3.4]{haenggi_book} have been used extensively over the last decade to model a variety of wireless network configurations over a 2D space \cite{azimi_cluster1, azimi_cluster2, azimi_cluster3, afshang_cluster1, afshang_cluster2, saha1,saha2}. They are particularly useful in capturing user hotspots that exhibit point clustering, which cannot be captured using a simpler homogeneous Poisson point process (PPP). Two of the specific PCPs that have been of particular interest in the applications are the Thomas cluster process (TCP) \cite[Definition 3.5]{haenggi_book} and the Mat{\'e}rn cluster process (MCP) \cite[Definition 3.6]{haenggi_book}. The formalism for establishing distance distributions of PCPs is well-known \cite{math_distance} and has been applied extensively to derive key distance distributions for PCPs \cite{azimi_cluster1,afshang_distance, afshang_distance2, gupta,pandey2021kth,pandey2020on}.

Very recently, 3D PCPs have been used to model and analyze biological nanonetworks by the authors \cite{azimi_bio}. The main idea is to model the locations of {\em molecular fusion centers} as a PPP that serves as the parent process for the {\em nanomachines} modeled as a PCP around the fusion centers. Since fusion centers have a non-zero size, it is more reasonable to model them as spheres instead of points, which essentially places exclusion zones (equivalently, {\em holes}) around the parent points of the PCPs where the nanomachines cannot exist. This results in a new variant of PCPs, which is the main topic of this paper. Our work in \cite{azimi_bio} also identified a specific structure of 3D TCPs that allowed us to express the distance distributions in 3D TCPs in remarkably simple closed form expressions (which was not possible in the simpler 2D case). This further inspired us to investigate the distributional properties of the aforementioned variant of PCPs in more detail in this paper. Before we describe our contributions, please note that the existence of exclusion zones in clustered processes can also be motivated from the perspective of wireless networks if one needs to ensure a certain minimum distance between the nodes of two different networks transmitting at the same frequency channel. When the holes are not placed on the cluster centers, \cite{afshang_hole} considered such a {\em cluster process model with holes} over a 2D space. This essentially generalized the idea of a {\em Poisson hole process (PHP)} \cite{hole} from a homogeneous PPP to the PCP with the common theme being that the underlying point process (PPP or PCP) is thinned by placing holes independently of the underlying point process. 


\begin{figure}	\vspace{0pt}	\centering	\includegraphics[width =3.0in]{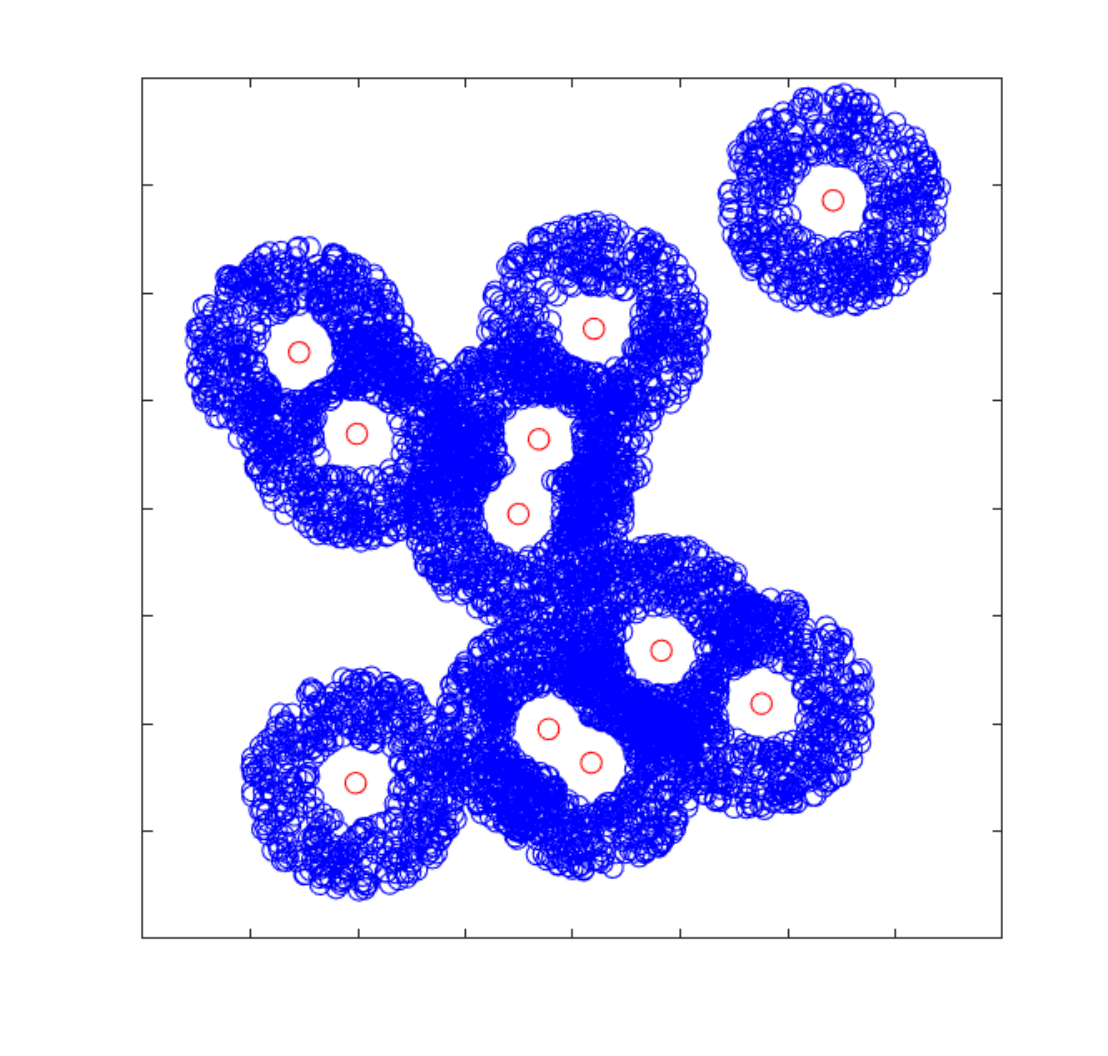} 	\vspace{-15pt}	\caption{An illustration of a 2D MCP-H, where the red points are the cluster centers and the blue points are the offspring points.}
\vspace{0pt}
\end{figure}

In this work, we first derive new distance distribution results for MCPs over 3D spaces, as the counterpart of similar results obtained for 3D TCPs in \cite{azimi_bio}. Then inspired by the recent applications of 3D PCPs to biological nanonetworks, we propose a new point process that we term the {\em Mat{\'e}rn cluster process with holes at the cluster centers (MCP-H)}, which is a variant of the MCP obtained by removing MCP points lying within a certain distance from any of the cluster centers (effectively placing exclusion zones or holes at each cluster center). While we present results for 3D, the definition and analytical approach can be easily extended to the $n$-dimensional space\footnote{Note that the 3D case is considered in this work because of its relevance to the underlying application of biological nanonetworks.}. We identify specific structures for both a 3D MCP and a MCP-H that provide remarkably simple expressions for the distribution of the distance between
an arbitrary point of a given cluster to the origin, conditioned on the location of that cluster. Using these distance distributions, we characterize the contact distance distributions and probability generating functionals (PGFL) of both MCP and MCP-H that are not only useful on their own but will also find use in the applications of these processes to diverse fields.

\section{Spatial Model}
Consider an MCP $\Phi_{\rm M}$ in $\mathbb{R}^3$, which is formally defined as a union of offspring points  that are located around parent points (i.e., cluster centers). The parent point process is a PPP $\Phi_\text{p}$ with intensity $\lambda_\text{p}$, and the offspring point processes (one per parent) are conditionally independent. The set of offspring points of $\mathbf{x} \in \Phi_\text{p}$ is a finite point process that is denoted by ${\cal N}_1^{{\mathbf{x}}}$, such that $\Phi_{\rm M} = \cup_{\mathbf{x}\in \Phi_{\rm p}}{\cal N}_1^{{\mathbf{x}}}$. The points in ${\cal N}_1^{{\mathbf{x}}}$ are uniformly distributed in the ball with the center $\mathbf{x}$ and radius $R$ represented by $\mathbf{b}(\mathbf{x}, R)$, i.e., ${\cal N}_1^{{\mathbf{x}}} \sim \text{Unif}\left\{\mathbf{b}(\mathbf{x}, R)\right\}$. Hence, the probability density function (PDF) of an element of this set being at a location $\mathbf{y} + \mathbf{x} \in \mathbb{R}^3$ is
\begin{align}
f_{{\mathbf{Y}}_1}(\mathbf{y}) = \left\{\begin{matrix}
\frac{3\|\mathbf{y}\|^2}{R^3} & \  \|\mathbf{y}\| \leq R,\\
0 & \ \text{o.w.}
\end{matrix}\right.
\end{align}
The number of points in ${\cal N}_1^\mathbf{x}, \forall \mathbf{x}$ is Poisson distributed with mean $M_1$. 

Along the same lines, the MCP-H is defined as $\Phi_{\rm {M-H}} = \cup_{\mathbf{x}\in \Phi_{\rm p}}{\cal N}_2^{{\mathbf{x}}}$, where the set ${\cal N}_2^{{\mathbf{x}}}$ is a thinned version of ${\cal N}_1^{{\mathbf{x}}}$. A point of ${\cal N}_1^{{\mathbf{x}}}$ can be thinned either by the exclusion zone at $\mathbf{x}$ or by the exclusion zones of any of the other points of $\Phi_\text{p}$. The resulting finite point process can be mathematically described by ${\cal N}_2^{{\mathbf{x}}} \sim {\cal N}_1^{{\mathbf{x}}} \backslash \cup_{\mathbf{z} \in \Phi_\text{p}} \mathbf{b}(\mathbf{z},r_0)$, or equivalently ${\cal N}_2^{{\mathbf{x}}} \sim \text{Unif}\left\{\mathbf{b}(\mathbf{x}, R)\right\} \backslash \cup_{\mathbf{z} \in \Phi_\text{p}} \mathbf{b}(\mathbf{z},r_0)$, where $r_0$ is the radius of the exclusion holes and is assumed to be the same for all the holes. The PDF of an element of ${\cal N}_2^{{\mathbf{x}}}$ being at a location $\mathbf{y} + \mathbf{x} \in \mathbb{R}^3$ is denoted by $f_{{\mathbf{Y}}_2}(\mathbf{y})$. A realization of the MCP-H is illustrated in Fig. 1.
The number of points in ${\cal N}_2^\mathbf{x}, \forall \mathbf{x}$ is Poisson distributed with mean $M_2$.

\section{Distance Distributions}
For both MCP and MCP-H, we first present the following results on the PDF of the distance of any (arbitrary) element in the set ${\cal N}_1^{{\mathbf{x}}}$ and ${\cal N}_2^{{\mathbf{x}}}$ of the cluster centered at $\mathbf{x}\in \Phi_{\text{p}}$ to the origin $\mathbf{o}$, respectively. Such PDF derivation is the key intermediate step required in PCPs for further analyses \cite{azimi_cluster1, azimi_cluster3, afshang_cluster1,afshang_cluster2,saha1,saha2,math_distance,afshang_distance,afshang_distance2,gupta,azimi_bio}. In order to illustrate this concretely, we will use these PDF results later in derivations of the contact distance distribution and PGFL in Section~\ref{sec:IV}. It is worth reiterating that even though the formalism for establishing distance distributions of point processes is well-known, our focus here is on demonstrating that these distributions admit a simple closed form solution in the case of 3D MCP, which is not even the case in the simpler 2D MCP. 

\begin{theorem} \label{thm:MCP}
	For the MCP and conditioned on $\|\mathbf{x}\|$, i.e., the distance of the parent point $\mathbf{x}$ from the origin, the PDF of the distances $d = \|\mathbf{y}+\mathbf{x}\|, \forall \mathbf{y}\in {\cal N}_1^\mathbf{x}$, for $\|\mathbf{x}\| \le R$ is
	\begin{align}
	f_d( r | \|\mathbf{x}\|) = \left\{\begin{matrix}
	\frac{ {3r^2}}{{ R^3}} & \  0\leq r<R-\|\mathbf{x}\|,\\
	\frac{3}{4}\frac{r(R-\|\mathbf{x}\|+r)(R+\|\mathbf{x}\|-r)}{R^3\|\mathbf{x}\|}  & \ R- \|\mathbf{x}\| \leq r<R+\|\mathbf{x}\|,\\ 0 & \ r\ge R+\|\mathbf{x}\|,
	\end{matrix}\right.
	\end{align}

	and for $\|\mathbf{x}\| > R$
	\begin{align}
	f_d( r | \|\mathbf{x}\|) =\left\{\begin{matrix}
	\frac{3}{4}\frac{r(R-\|\mathbf{x}\|+r)(R+\|\mathbf{x}\|-r)}{R^3\|\mathbf{x}\|}  & \ \|\mathbf{x}\|-R \leq r<R+\|\mathbf{x}\|,\\
	0 &\ 0\leq r<\|\mathbf{x}\| -R, \ r\ge R+\|\mathbf{x}\|.
	\end{matrix}\right.
	\end{align}	
	
\end{theorem}
\begin{IEEEproof}
	See Appendix A. 
\end{IEEEproof}

The exact derivation for the distance distributions of MCP-H is complicated. This is because of the complicated characterization of the region $\mathbf{b}(\mathbf{x}, R) \backslash \cup_{\mathbf{z} \in \Phi_\text{p}} \mathbf{b}(\mathbf{z},r_0)$ that defines the support for the offspring points in the cluster $\mathbf{x}$. Therefore, our first objective is to bound this distance. This bound is constructed by bounding the aforementioned region as $\mathbf{b}(\mathbf{x}, R) \backslash \cup_{\mathbf{z} \in \Phi_\text{p}} \mathbf{b}(\mathbf{z},r_0) \subset \mathbf{b}(\mathbf{x}, R) \backslash \mathbf{b}(\mathbf{x},r_0)$. In other words, we retain the most important hole (the one centered at $\mathbf{x}$ or the {\em self hole}) and ignore effect of the other holes on the point process ${\cal N}_2^{{\mathbf{x}}}$. Since less points are thinned than necessary, this underestimates the distance and hence results in an upper bound on the cumulative distribution function (CDF) whose corresponding PDF is given in the next Theorem. This bounding approach is inspired by a similar thought process that was used by the authors to bound distances in the PHP in \cite{hole}.
\begin{theorem}  \label{thm:MCP-H-UpperBound}
	For the MCP-H and conditioned on $\|\mathbf{x}\|$, the PDF of the distances $d = \|\mathbf{y}+\mathbf{x}\|, \forall \mathbf{y}\in {\cal N}_2^\mathbf{x}$, corresponding to the upper bound on the CDF mentioned above is given in the following cases.
	
	\textbf{Case 1:} If $0 \le \|\mathbf{x}\| < \min\left\{r_0,\frac{R-r_0}{2}\right\}$, then
	\begin{align}
	f_d^\text{u}( r | \|\mathbf{x}\|) = \left\{\begin{matrix}
	0  &\ 0\leq r<r_0 -\|\mathbf{x}\|,\ r\ge R+\|\mathbf{x}\|,\\
	\frac{3r^2-\frac{3r}{4\|\mathbf{x}\|}(r_0-\|\mathbf{x}\|+r)(r_0+\|\mathbf{x}\|-r)}{R^3-r_0^3} &\  r_0 -\|\mathbf{x}\| \leq r<r_0+\|\mathbf{x}\|,\\
	\frac{3r^2}{{R^3-r_0^3}} &\  \|\mathbf{x}\|+r_0 \leq r<R-\|\mathbf{x}\|,\\
	\frac{3r(R-\|\mathbf{x}\|+r)(R+\|\mathbf{x}\|-r)}{4\|\mathbf{x}\|(R^3-r_0^3)} &\  R-\|\mathbf{x}\| \leq r<R+\|\mathbf{x}\|.
	\end{matrix}\right.
	\end{align}
	\textbf{Case 2:} If $\min\left\{r_0,\frac{R-r_0}{2}\right\} \le \|\mathbf{x}\| < r_0$, then
	\begin{align}
	f_d^\text{u}( r | \|\mathbf{x}\|) =  \left\{\begin{matrix}
	0  &\ 0\leq r<r_0 -\|\mathbf{x}\|, \ r\ge R+\|\mathbf{x}\|,\\
	\frac{3r^2-\frac{3r}{4\|\mathbf{x}\|}(r_0-\|\mathbf{x}\|+r)(r_0+\|\mathbf{x}\|-r)}{R^3-r_0^3} &\  r_0 -\|\mathbf{x}\| \leq r<R-\|\mathbf{x}\|,\\
	\frac{3}{4}\frac{r(R^2-r_0^2)}{(R^3-r_0^3)\|\mathbf{x}\|} &\  R-\|\mathbf{x}\| \leq r<r_0+\|\mathbf{x}\|,\\
	\frac{3r(R-\|\mathbf{x}\|+r)(R+\|\mathbf{x}\|-r)}{4\|\mathbf{x}\|(R^3-r_0^3)} &\  r_0+\|\mathbf{x}\| \leq r<R+\|\mathbf{x}\|.
	\end{matrix}\right.
	\end{align}
	\textbf{Case 3:} If $r_0 \le \|\mathbf{x}\| < \max\left\{r_0,\frac{R-r_0}{2}\right\}$, then
	\begin{align}
	f_d^\text{u}( r | \|\mathbf{x}\|) = \left\{\begin{matrix}
	\frac{ {3r^2}}{{ R^3-r_0^3}}  & \ 0\leq r<\|\mathbf{x}\|-r_0,\\&\  \|\mathbf{x}\|+r_0 \leq r<R-\|\mathbf{x}\|,\\
	\frac{3r^2-\frac{3r}{4\|\mathbf{x}\|}(r_0-\|\mathbf{x}\|+r)(r_0+\|\mathbf{x}\|-r)}{R^3-r_0^3} &\ \|\mathbf{x}\|-r_0 \leq r<\|\mathbf{x}\|+r_0,
	\\
	\frac{3r(R-\|\mathbf{x}\|+r)(R+\|\mathbf{x}\|-r)}{4\|\mathbf{x}\|(R^3-r_0^3)} &\ R-\|\mathbf{x}\| \leq r<R+\|\mathbf{x}\|,\\
	0 & r\ge R+\|\mathbf{x}\|.
	\end{matrix}\right.
	\end{align}	
	\textbf{Case 4:} If $\max\left\{r_0,\frac{R-r_0}{2}\right\} \le \|\mathbf{x}\| < \frac{R+r_0}{2}$, then	
	\begin{align}
	f_d^\text{u}( r | \|\mathbf{x}\|) = \left\{\begin{matrix}
	\frac{ {3r^2}}{{ R^3-r_0^3}}& \ 0 \leq r<\|\mathbf{x}\|-r_0,\\
	\frac{3r^2-\frac{3r}{4\|\mathbf{x}\|}(r_0-\|\mathbf{x}\|+r)(r_0+\|\mathbf{x}\|-r)}{R^3-r_0^3} &\ \|\mathbf{x}\|-r_0 \leq r<R-\|\mathbf{x}\|, \\
	\frac{3}{4}\frac{r(R^2-r_0^2)}{(R^3-r_0^3)\|\mathbf{x}\|}& \ R-\|\mathbf{x}\| \leq r<\|\mathbf{x}\|+r_0,
	\\
	\frac{3r(R-\|\mathbf{x}\|+r)(R+\|\mathbf{x}\|-r)}{4\|\mathbf{x}\|(R^3-r_0^3)} &\ \|\mathbf{x}\|+r_0 \leq r<R+\|\mathbf{x}\|,\\
	0 & r\ge R+\|\mathbf{x}\|.
	\end{matrix}\right.
	\end{align}	
	\textbf{Case 5:} If $\frac{R+r_0}{2} \le \|\mathbf{x}\| <R$, then
	\begin{align}
	f_d^\text{u}( r | \|\mathbf{x}\|) = \left\{\begin{matrix}
	\frac{ {3r^2}}{{ R^3-r_0^3}}& \ 0 \leq r<R-\|\mathbf{x}\|,\\
	\frac{3r(R-\|\mathbf{x}\|+r)(R+\|\mathbf{x}\|-r)}{4\|\mathbf{x}\|(R^3-r_0^3)} &\ R-\|\mathbf{x}\| \leq r<\|\mathbf{x}\|-r_0,\\&\
	\|\mathbf{x}\|+r_0 \leq r<R+\|\mathbf{x}\|,
	\\
	\frac{3}{4}\frac{r(R^2-r_0^2)}{(R^3-r_0^3)\|\mathbf{x}\|}& \ \|\mathbf{x}\|-r_0 \leq r<\|\mathbf{x}\|+r_0,
	\\
	0 & r\ge R+\|\mathbf{x}\|.
	\end{matrix}\right.
	\end{align}		
	\textbf{Case 6:} If $\|\mathbf{x}\| \ge R$, then
	\begin{align}
	f_d^\text{u}( r | \|\mathbf{x}\|) = \left\{\begin{matrix}
	0& \ 0 \leq r<\|\mathbf{x}\|-R,\
	r\ge R+\|\mathbf{x}\|,
	\\
	\frac{3r(R-\|\mathbf{x}\|+r)(R+\|\mathbf{x}\|-r)}{4\|\mathbf{x}\|(R^3-r_0^3)} &\ \|\mathbf{x}\|-R \leq r<\|\mathbf{x}\|-r_0,\\&\
	\|\mathbf{x}\|+r_0 \leq r<R+\|\mathbf{x}\|,
	\\
	\frac{3}{4}\frac{r(R^2-r_0^2)}{(R^3-r_0^3)\|\mathbf{x}\|}& \ \|\mathbf{x}\|-r_0 \leq r<\|\mathbf{x}\|+r_0.
	\end{matrix}\right.
	\end{align}		
	
\end{theorem}
\begin{IEEEproof}
	See Appendix B. 
\end{IEEEproof}

In Theorem~\ref{thm:MCP-H-UpperBound}, the thinning effect of the other clusters on the locations of the offspring points in a cluster is ignored. However, the overlapping holes of other clusters can reduce the support of the offspring point process (i.e., the region in which the offspring points are distributed). One can think of many ways of approximating the effect of these other holes. For instance, we can replace $\mathbf{b}(\mathbf{x}, R) \backslash \cup_{\mathbf{z} \in \Phi_\text{p}} \mathbf{b}(\mathbf{z},r_0)$ with an effective region that approximately incorporates the effect of the other holes. The distance distribution result can then be obtained by repeating the proof of Appendix B with a slight modification in the intersection volume. As a simple example, from the intersection volume $\mathbf{b}(\mathbf{o}, r) \cap \mathbf{b}(\mathbf{x}, R)\backslash \mathbf{b}(\mathbf{x}, r_0)$, we can reduce an effective volume that is equal to the average number of holes from all clusters inside $\mathbf{b}(\mathbf{o}, r) \cap \mathbf{b}(\mathbf{x}, R)\backslash \mathbf{b}(\mathbf{x}, r_0)$ multiplied by the volume of each hole. This leads to a scaling factor of $1- \frac{4}{3} \pi \lambda_\text{p}r_0^3$ on $f_d^\text{u}(r|\|\mathbf{x}\|)$. This can be tightened further by accounting for the overlaps in the holes, along the lines of how we handled such overlaps in the PHP in \cite{hole}. Due to strict space limitations, we will not be able to explore such approximations in this article.

\section{Point Process Properties} \label{sec:IV}
In this section, we will use the distance distributions derived in the previous section to characterize two key properties of the 3D MCP and MCP-H. The first property is the contact distance distribution. In a stationary setting, this is simply the distribution of the distance between the origin and the closest point of the point process to the origin. This appears frequently in many applications, such as the distance between a user and its closest wireless tower in a wireless network. The second property is the PGFL, which allows us to deal with products over point processes. It plays a central role in the analysis of interference distribution in wireless networks. 

\subsection{Contact Distance Distribution}
\begin{definition}
	(Contact distance distribution). The contact distance of a stationary point process $\Phi$ is defined as the distance from the origin to its nearest point. The distribution of this distance is termed the contact distance distribution of the point process $\Phi$ and is defined as
	\begin{align}
	\label{contact}
	F_\text{CD}(r) = \mathbb{P}\left\{\|\Phi\|\leq r \right\} = \mathbb{P}\left\{\Phi(\mathbf{b}(\mathbf{o},r))>0 \right\},
	\end{align}
	where $\Phi(\mathbf{b}(\mathbf{o},r))$ denotes
	the number of points within a ball of radius $r$ centered at $\mathbf{o}$.
\end{definition}
To derive the contact distance distribution as in \eqref{contact}, let us start with the derivation of the probability generating
function (PGF) of the number of points within $\mathbf{b}(\mathbf{o},r)$ as
\begin{align}
\label{Gt1}
&G_N(\theta) = \mathbb{E}\left\{\theta^N\right\} = \mathbb{E}\left\{\theta^{\sum_{\mathbf{x} \in \Phi_\text{p}}\sum_{\mathbf{y} \in {\cal N}_i^\mathbf{x}}\mathds{1}(\|\mathbf{x}+\mathbf{y}\|<r)}\right\} =
\mathbb{E}\left\{\mathop \prod \limits_{\mathbf{x} \in \Phi_\text{p}}\mathop \prod \limits_{\mathbf{y} \in {\cal N}_i^{\mathbf{x}}} \theta^{\mathds{1}(\|\mathbf{x}+\mathbf{y}\|<r)}\right\} \nonumber\\&=
\mathbb{E}\Biggl\{\mathop \prod \limits_{\mathbf{x} \in \Phi_\text{p}}\exp\biggl(-M_i \int_{\mathbf{b}(\mathbf{x},R)}\left(1-\theta^{\mathds{1}(\|\mathbf{x}+\mathbf{y}\|<r)}\right)f_{\mathbf{Y}_i}(\mathbf{y})\mathrm{d}\mathbf{y}\biggr)\Biggr\}, \ i \in \left\{1, 2\right\},
\end{align}
where the last step follows from the PGF of a Poisson random variable with the fact that the points in ${\cal N}_i^{\mathbf{x}}$ are i.i.d. with PDF $f_{\mathbf{Y}_i}(\mathbf{y})$. The index $i=1$ is for the MCP and the index $i=2$ is for the MCP-H. Then, \eqref{Gt1} for the MCP is equal to and for the MCP-H is approximately equal to 
\begin{align}
\label{app}
\exp\Biggl(-4\pi \lambda_\text{p}\int_{0}^{\infty}\biggl(1-\exp\Bigl(-M_i\int_{0}^{r}(1-\theta)f_d(u|v) \mathrm{d}u\Bigr)\biggr)v^2 \mathrm{d}v\Biggr), \ i \in \left\{1, 2\right\},
\end{align}
which follows from the PGFL of PPPs \cite{haenggi_book} and converting Cartesian to polar
coordinates with the distance distribution $f_d(u|v)$ in Theorems 1-2. Equation \eqref{app} is not exact for MCP-H because of the bound in Theorem 2.

For the MCP, we can obtain \eqref{Gt1} from $f_d(u|v)$ given in Theorem 1 as 
\begin{align}
\label{Gt2}
&G_{N,\text{MCP}}(\theta) = \exp\Biggl(-4\pi \lambda_\text{p}\int_{0}^{R}\biggl(1-\exp\Bigl(-M_1(1-\theta)\Bigl(\int_{0}^{\min\left\{r,R-v\right\}}\frac{ {3u^2}}{{ R^3}}\mathrm{d}u\nonumber\\
&+\int_{\min\left\{r,R-v\right\}}^{\min\left\{r,R+v\right\}}\frac{3}{4}\frac{u(R-v+u)(R+v-u)}{R^3v}\mathrm{d}u\Bigr)\Bigr)\biggr)v^2 \mathrm{d}v-\nonumber\\
&4\pi \lambda_\text{p}\int_{R}^{\infty}\biggl(1-\exp\Bigl(-M_1(1-\theta)\int_{\min\left\{r,v-R\right\}}^{\min\left\{r,v+R\right\}}\frac{3}{4}
\frac{u(R-v+u)(R+v-u)}{R^3v}\mathrm{d}u\Bigr)\biggr) v^2\mathrm{d}v\Biggr)\nonumber\end{align}
\begin{align}
&=\exp\Biggl(-4\pi \lambda_\text{p}\int_{0}^{R}\biggl(1-\exp\Bigl(-M_1(1-\theta)\Bigl(\frac{\min \left\{r,(R-v)\right\}^3}{R^3}+\frac{3}{4R^3v}\times\nonumber\\&\Bigl(\frac{(R^2-v^2)(\min \left\{r,(R+v)\right\}^2-\min \left\{r,(R-v)\right\}^2)}{2}+\nonumber\\
&\frac{2v(\min \left\{r,(R+v)\right\}^3-\min \left\{r,(R-v)\right\}^3)}{3}-\frac{(\min \left\{r,(R+v)\right\}^4-\min \left\{r,(R-v)\right\}^4)}{4}\nonumber\\&\Bigr)\Bigr)\Bigr)\biggr) v^2\mathrm{d}v-4\pi \lambda_\text{p}\int_{R}^{\infty}\biggl(1-\exp\Bigl(-M_1(1-\theta)\frac{3}{4R^3v}\Bigl((R^2-v^2)\times\nonumber\\&\frac{\min \left\{r,(R+v)\right\}^2-\min \left\{r,(v-R)\right\}^2}{2}+\frac{2v(\min \left\{r,(R+v)\right\}^3-\min \left\{r,(v-R)\right\}^3)}{3}\nonumber\\&-\frac{(\min \left\{r,(R+v)\right\}^4-\min \left\{r,(v-R)\right\}^4)}{4}\Bigr)\Bigr)\biggr) v^2\mathrm{d}v\Biggr).
\end{align}
This result for the 3D MCP is in closed-form and has completely different structure from the 2D MCP result in \cite{afshang_distance2}, which includes multiple integrals in its simplest form. 

Now, defining
\begin{align}
P(a,b,c,d) &= \frac{3}{4cv}\Biggl(\frac{(d^2-v^2)(\min \left\{r,a\right\}^2-\min \left\{r,b\right\}^2)}{2}+\nonumber\\&\frac{2v(\min \left\{r,a\right\}^3-\min \left\{r,b\right\}^3)}{3}-\frac{(\min \left\{r,a\right\}^4-\min \left\{r,b\right\}^4)}{4}\Biggr),
\end{align}
and following a similar approach as in \eqref{Gt2}, we obtain \eqref{Gt1} for the MCP-H, replacing $f_d(u|v)$ in Theorem 2, as
\begin{align}
&G_{N, \text{MCP-H}}(\theta) \approx \exp\Biggl(-4\pi \lambda_\text{p}\int_{0}^{\min\left\{r_0,\frac{R-r_0}{2}\right\}}\biggl(1-\exp\Bigl(-M_2(1-\theta)A_1(r)\Bigr)\biggr) v^2\mathrm{d}v\nonumber\\&-4\pi \lambda_\text{p}\int_{\min\left\{r_0,\frac{R-r_0}{2}\right\}}^{r_0}\biggl(1-\exp\Bigl(-M_2(1-\theta)B_1(r)\Bigr)\biggr) v^2\mathrm{d}v-4\pi \lambda_\text{p}\int_{r_0}^{\max\left\{r_0,\frac{R-r_0}{2}\right\}}\nonumber\\&\biggl(1-\exp\Bigl(-M_2(1-\theta)C_1(r)\Bigr)\biggr) v^2\mathrm{d}v-4\pi \lambda_\text{p}\int_{\max\left\{r_0,\frac{R-r_0}{2}\right\}}^{\frac{R+r_0}{2}}\biggl(1-\exp\Bigl(-M_2(1-\theta)\nonumber\\
& D_1(r)\Bigr)\biggr) v^2\mathrm{d}v-4\pi \lambda_\text{p}\int_{\frac{R+r_0}{2}}^{R}\biggl(1-\exp\Bigl(-M_2(1-\theta)E_1(r)\Bigr)\biggr) v^2\mathrm{d}v\nonumber\\
&-4\pi \lambda_\text{p}\int_{R}^{\infty}\biggl(1-\exp\Bigl(-M_2(1-\theta)F_1(r)\Bigr)\biggr) v^2\mathrm{d}v\Biggr),
\end{align}
where
\begin{align}
A_1(r) &= \frac{\min \left\{r,(R-v)\right\}^3}{R^3-r_0^3} -\frac{\min \left\{r,(r_0-v)\right\}^3}{R^3-r_0^3} -P(v+r_0,r_0-v,R^3-r_0^3,r_0)\nonumber\\&+P(R+v,R-v,R^3-r_0^3,R),
\end{align}
\begin{align}
&B_1(r) = \frac{\min \left\{r,(R-v)\right\}^3}{R^3-r_0^3} -\frac{\min \left\{r,(r_0-v)\right\}^3}{R^3-r_0^3} -P(R-v,r_0-v,R^3-r_0^3,r_0)+\nonumber\\&P(R+v,v+r_0,R^3-r_0^3,R)+\frac{3}{8}\frac{(R^2-r_0^2)(\min \left\{(v+r_0),r\right\}^2-\min \left\{(R-v),r\right\}^2)}{(R^3-r_0^3)v},
\end{align}
\vspace{0pt}
\begin{align}
C_1(r) &= \frac{\min \left\{r,(R-v)\right\}^3}{R^3-r_0^3} -P(v+r_0,v-r_0,R^3-r_0^3,r_0)\nonumber\\&+P(R+v,R-v,R^3-r_0^3,R),
\end{align}
\vspace{0pt}
\begin{align}
&D_1(r) = \frac{\min \left\{r,(R-v)\right\}^3}{R^3-r_0^3}-P(R-v,v-r_0,R^3-r_0^3,r_0)+\nonumber\\&P(R+v,v+r_0,R^3-r_0^3,R)+\frac{3}{8}\frac{(R^2-r_0^2)(\min \left\{(v+r_0),r\right\}^2-\min \left\{(R-v),r\right\}^2)}{(R^3-r_0^3)v},
\end{align}
\begin{align}
&E_1(r) = \frac{\min \left\{r,(R-v)\right\}^3}{R^3-r_0^3}+P(v-r_0,R-v,R^3-r_0^3,R)+\nonumber\\&P(R+v,v+r_0,R^3-r_0^3,R)+\frac{3}{8}\frac{(R^2-r_0^2)(\min\left\{(v+r_0),r\right\}^2-\min\left\{(v-r_0),r\right\}^2)}{(R^3-r_0^3)v},
\end{align}
\begin{align}
&F_1(r) = P(v-r_0,v-R,R^3-r_0^3,R)+P(R+v,v+r_0,R^3-r_0^3,R)\nonumber\\&+\frac{3}{8}\frac{(R^2-r_0^2)(\min\left\{(v+r_0),r\right\}^2-\min\left\{(v-r_0),r\right\}^2)}{(R^3-r_0^3)v}.
\end{align}

Finally, the contact distance distribution can be obtained as
\begin{align}
\label{FCD}
F_{\text{CD},j}(r) = 1-\mathbb{P}(N = 0) = 1- G_{N,j}(0), \ j=\left\{\text{MCP},\text{MCP-H}\right\}.
\end{align}
It is notable that all the results in \eqref{Gt2}-\eqref{FCD} are in closed-forms. 

\begin{figure}[tb!]	\vspace{0pt}	\centering	\includegraphics[width =3.6in]{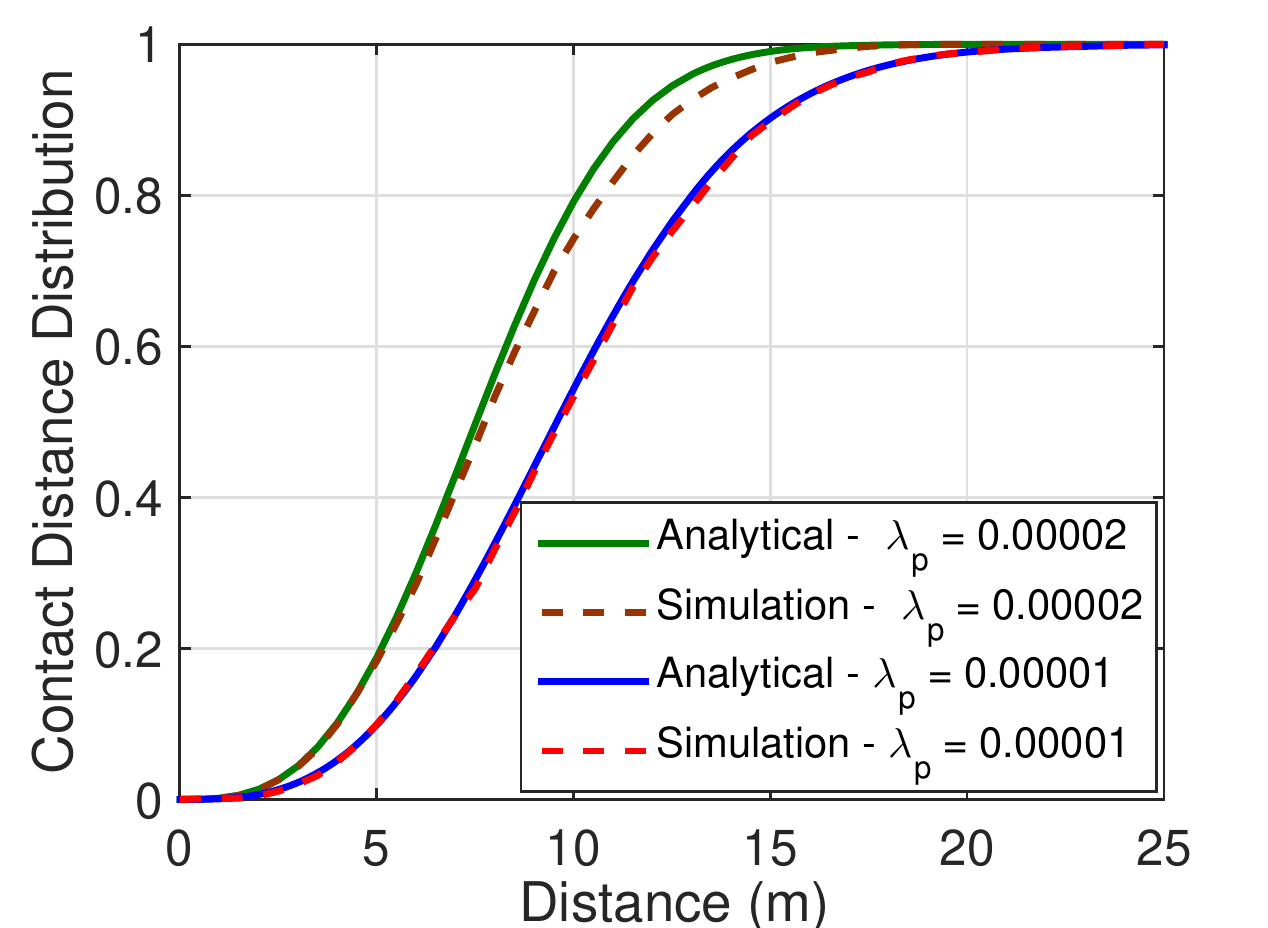} 	\vspace{0pt}	\caption{The contact distance distribution for the MCP-H.}
	\vspace{-10pt}
	\label{fig:MCP-H}
\end{figure}

In Fig.~\ref{fig:MCP-H}, we plot the contact distance distributions for the MCP-H, for different $\lambda_\text{p}$ when $R = 50 \ \text{m}$, $r_0 = 15 \ \text{m}$, and $M_2 = 20$. We note that the bound derived for MCP-H is tight even for a relatively large value of $\lambda_\text{p}$ with respect to the system setup. In order to put this comparison in context, note that for $\lambda_\text{p} = 0.00002$, the average number of holes of the other clusters that overlap with the reference cluster is $\frac{4}{3}\pi R^3 \lambda_\text{p} = 10.5$. Despite that, the bound is remarkably tight in this regime. If we keep increasing $\lambda_\text{p}$ to very high values, the bound will of course start becoming loose. Even though it is not pronounced, this trend is visible in Fig.~\ref{fig:MCP-H} as we move from $\lambda_\text{p} = 0.00001$ to $0.00002$. 

\subsection{Probability Generating Functional}
\begin{definition}(PGFL).
	The PGFL of a point process $\Phi$ is defined as
	\begin{align}
	G(v(\mathbf{y})) = \mathbb{E}\left\{\prod_{\mathbf{y} \in \Phi}v(\mathbf{y})\right\},
	\end{align}
	where the function $v(\mathbf{y}): \mathbb{R}^3 \to [0,1]$ is measurable.
\end{definition}
According to the PGFL of PCPs \cite[Corollary 4.13]{haenggi_book} and replacing $f_d(\|\mathbf{y}\| | \|\mathbf{z}\|)$ from Theorems 1-2,  the PGFL for the MCP and MCP-H can be evaluated exactly and approximately, respectively, as
\begin{align}
\label{pgfl}
&G_j(v(\mathbf{y}))  \approxeq \exp \biggl(-\lambda_\text{p} \int_{\mathbb{R}^3}^{}\biggl(1-\exp \biggl(-M_i \biggl(1-\int_{\mathbb{R}^3}v(\mathbf{y})f_d(\|\mathbf{y}\| | \|\mathbf{z}\|)\mathrm{d}\mathbf{y}\biggr)\biggr)\biggr)\mathrm{d}\mathbf{z}\biggr), \nonumber\\& \hspace{170pt}(i,j) = \left\{(1,\text{MCP}),(2,\text{MCP-H})\right\}.
\end{align}
The result for the MCP-H
is not exact because of the bound in Theorem 2. Due to the very simple forms of $f_d(u|v)$ in Theorems~\ref{thm:MCP} and \ref{thm:MCP-H-UpperBound}, \eqref{pgfl} can be efficiently evaluated for different functions $v(\mathbf{y})$. Note again that the result is exact for MCP since the distance distribution in Theorem~\ref{thm:MCP} was exact.

As mentioned already, PGFLs of point processes appear frequently in applications. For instance, it has been used extensively for deriving the Laplace transform of interference distribution in a variety of wireless network settings. Interested readers are advised to refer to~\cite{Cellular-SG-Book} for specific examples related to the wireless cellular networks.

\section{Conclusions}
This paper studied a novel variant of an MCP in which the points located within a certain distance from the cluster center are removed (or thinned). We termed this variant the MCP-H. Specifically focusing on the 3D setup, we characterized basic distance distributions for both MCP and MCP-H, which admitted remarkably simple closed forms (which were not possible even in the simpler 2D setting). Using these distance distributions, we characterized the PGFL and contact distance distributions of both the processes. The proposed point process has numerous applications ranging from nanonetworks to large-scale wireless networks.

\appendices
\section{Proof of Theorem 1}
The distance from the origin $\mathbf{o}$ to a randomly
chosen offspring point $\mathbf{y}$ with parent $\mathbf{x}$, i.e., $\|\mathbf{x}+\mathbf{y}\| = d$, is less than $r$ if and only if the point is located within the intersection $\mathbf{b}(\mathbf{o}, r) \cap \mathbf{b}(\mathbf{x}, R)$.
Thus, as each point is distributed independently and
uniformly within $\mathbf{b}(\mathbf{x}, R)$, according to different forms of the intersection volume, we have the following cases for the CDF of $d$.

\textbf{Case 1:} If $\|\mathbf{x}\| \le R$, then
\begin{align}
\label{appa1}
{\mathbb{P}}\left( {d \leq r} \right) = \left\{\begin{matrix}
\frac{{r^3}}{{ R^3}} & \ 0\leq r<R-\|\mathbf{x}\|,\\
\frac{{\cal V}_{\|\mathbf{x}\|}(r,R)}{{ \frac{4}{3}\pi R^3}} &\ R-\|\mathbf{x}\| \leq r<R+\|\mathbf{x}\|,
\\
1 & r\ge R+\|\mathbf{x}\|,
\end{matrix}\right.
\end{align}	
where ${\cal V}_{\|\mathbf{x}\|}(r,R)$ is the intersection volume of $\mathbf{b}(\mathbf{o}, r) \cap \mathbf{b}(\mathbf{x}, R)$ for $|R-\|\mathbf{x}\|| \le r \le R+\|\mathbf{x}\|$ given as \cite{wolfram}
\begin{align}
&{\cal V}_{\|\mathbf{x}\|}(r,R) =\frac{\pi (R+r-\|\mathbf{x}\|)^2}{12\|\mathbf{x}\|}   {(\|\mathbf{x}\|^2+2\|\mathbf{x}\|r-3r^2+2\|\mathbf{x}\|R+6rR-3R^2)}.
\end{align}

\textbf{Case 2:} If $\|\mathbf{x}\| > R$, then
\begin{align}
\label{appa2}
{\mathbb{P}}\left( {d \leq r} \right) = \left\{\begin{matrix}
0 & \ 0\leq r<\|\mathbf{x}\|-R,\\
\frac{{\cal V}_{\|\mathbf{x}\|}(r,R)}{{ \frac{4}{3}\pi R^3}} &\ \|\mathbf{x}\|-R \leq r<R+\|\mathbf{x}\|,
\\
1 & r\ge R+\|\mathbf{x}\|.
\end{matrix}\right.
\end{align}

Thus, the PDF is obtained by taking derivative of ${\mathbb{P}}\left( {d \leq r} \right)$ for both cases in \eqref{appa1} and \eqref{appa2}, where we have
\begin{align}
\frac{\mathrm{d}{\cal V}_{\|\mathbf{x}\|}(r,R)}{\mathrm{d}r} = \frac{\pi r}{\|\mathbf{x}\|}(R+r-\|\mathbf{x}\|)(R-r+\|\mathbf{x}\|).
\end{align}

\section{Proof of Theorem 2} \label{appendix:B}
The distance $\|\mathbf{x}+\mathbf{y}\| = d$ is less than $r$ if and only if the point is located within the intersection $\mathbf{b}(\mathbf{o}, r) \cap \mathbf{b}(\mathbf{x}, R)$ and outside $ \mathbf{b}(\mathbf{x}, r_0)$. In Fig. 3, an example of the intersection is shown.
Thus, according to different forms of the volume of $\mathbf{b}(\mathbf{o}, r) \cap \mathbf{b}(\mathbf{x}, R)\backslash \mathbf{b}(\mathbf{x}, r_0)$, we have the following cases for the CDF of $d$. 
\begin{figure}[tb!]	\vspace{0pt}	\centering	\includegraphics[width =1.95in]{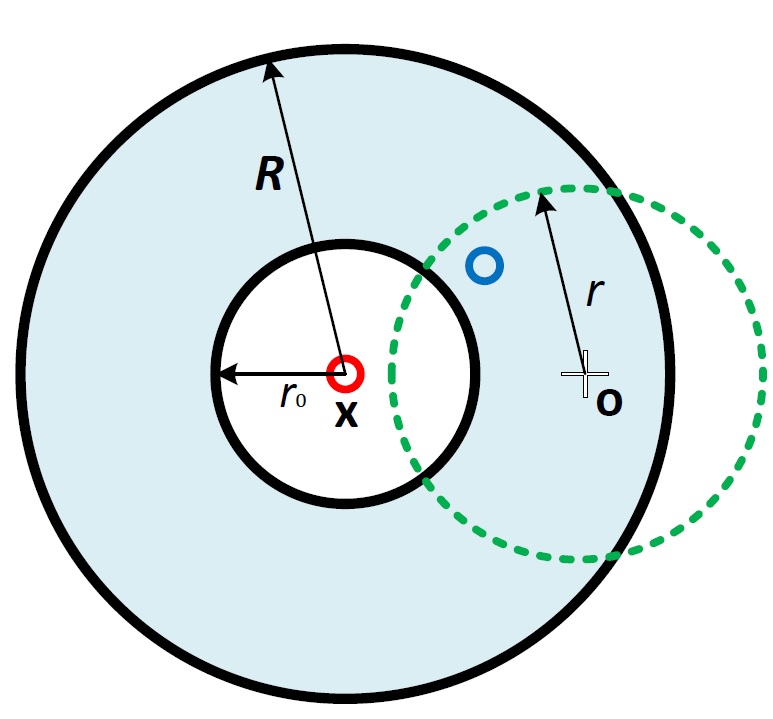} 	\vspace{0pt}	\caption{ An illustration of the intersection.}
	\vspace{-10pt}
\end{figure}

\textbf{Case 1:} If $0 \le \|\mathbf{x}\| < \min\left\{r_0,\frac{R-r_0}{2}\right\}$, then
\begin{align}
{\mathbb{P}}\left( {d \leq r} \right) = \left\{\begin{matrix}
0  &\ 0\leq r<r_0 -\|\mathbf{x}\|,\\
\frac{\frac{4}{3}\pi r^3 -{\cal V}_{\|\mathbf{x}\|}(r,r_0)}{{ \frac{4}{3}\pi (R^3-r_0^3)}} &\  r_0 -\|\mathbf{x}\| \leq r<r_0+\|\mathbf{x}\|,\\
\frac{r^3- r_0^3}{{R^3-r_0^3}} &\  \|\mathbf{x}\|+r_0 \leq r<R-\|\mathbf{x}\|,\\
\frac{{\cal V}_{\|\mathbf{x}\|}(r,R)-\frac{4}{3}\pi r_0^3}{{ \frac{4}{3}\pi (R^3-r_0^3)}} &\  R-\|\mathbf{x}\| \leq r<R+\|\mathbf{x}\|,
\\
1 & r\ge R+\|\mathbf{x}\|.
\end{matrix}\right.
\end{align}

\textbf{Case 2:} If $\min\left\{r_0,\frac{R-r_0}{2}\right\} \le \|\mathbf{x}\| < r_0$, then
\begin{align}
{\mathbb{P}}\left( {d \leq r} \right) = \left\{\begin{matrix}
0  &\ 0\leq r<r_0 -\|\mathbf{x}\|,\\
\frac{\frac{4}{3}\pi r^3 -{\cal V}_{\|\mathbf{x}\|}(r,r_0)}{{ \frac{4}{3}\pi (R^3-r_0^3)}} &\  r_0 -\|\mathbf{x}\| \leq r<R-\|\mathbf{x}\|,\\
\frac{{\cal V}_{\|\mathbf{x}\|}(r,R)-{\cal V}_{\|\mathbf{x}\|}(r,r_0)}{{ \frac{4}{3}\pi (R^3-r_0^3)}} &\  R-\|\mathbf{x}\| \leq r<r_0+\|\mathbf{x}\|,\\
\frac{{\cal V}_{\|\mathbf{x}\|}(r,R)-\frac{4}{3}\pi r_0^3}{{ \frac{4}{3}\pi (R^3-r_0^3)}} &\  r_0+\|\mathbf{x}\| \leq r<R+\|\mathbf{x}\|,
\\
1 & r\ge R+\|\mathbf{x}\|.
\end{matrix}\right.
\end{align}

\textbf{Case 3:} If $r_0 \le \|\mathbf{x}\| < \max\left\{r_0,\frac{R-r_0}{2}\right\}$, then
\begin{align}
{\mathbb{P}}\left( {d \leq r} \right) = \left\{\begin{matrix}
\frac{ {r^3} }{{ R^3-r_0^3}}  &\ 0\leq r<\|\mathbf{x}\|-r_0,\\
\frac{\frac{4}{3}\pi r^3 -{\cal V}_{\|\mathbf{x}\|}(r,r_0)}{{ \frac{4}{3}\pi (R^3-r_0^3)}} &\  \|\mathbf{x}\|-r_0 \leq r<\|\mathbf{x}\|+r_0,\\
\frac{r^3- r_0^3}{{R^3-r_0^3}} &\  \|\mathbf{x}\|+r_0 \leq r<R-\|\mathbf{x}\|,\\
\frac{{\cal V}_{\|\mathbf{x}\|}(r,R)-\frac{4}{3}\pi r_0^3}{{ \frac{4}{3}\pi (R^3-r_0^3)}} &\  R-\|\mathbf{x}\| \leq r<R+\|\mathbf{x}\|,
\\
1 & r\ge R+\|\mathbf{x}\|.
\end{matrix}\right.
\end{align}

\textbf{Case 4:} If $\max\left\{r_0,\frac{R-r_0}{2}\right\} \le \|\mathbf{x}\| < \frac{R+r_0}{2}$, then
\begin{align}
{\mathbb{P}}\left( {d \leq r} \right) = \left\{\begin{matrix}
\frac{ {r^3} }{{ R^3-r_0^3}}  &\ 0\leq r<\|\mathbf{x}\|-r_0,\\
\frac{\frac{4}{3}\pi r^3 -{\cal V}_{\|\mathbf{x}\|}(r,r_0)}{{ \frac{4}{3}\pi (R^3-r_0^3)}} &\  \|\mathbf{x}\|-r_0 \leq r<R-\|\mathbf{x}\|,\\
\frac{{\cal V}_{\|\mathbf{x}\|}(r,R)-{\cal V}_{\|\mathbf{x}\|}(r,r_0)}{{R^3-r_0^3}} &\  R-\|\mathbf{x}\| \leq r<\|\mathbf{x}\|+r_0,\\
\frac{{\cal V}_{\|\mathbf{x}\|}(r,R)-\frac{4}{3}\pi r_0^3}{{ \frac{4}{3}\pi (R^3-r_0^3)}} &\  \|\mathbf{x}\|+r_0 \leq r<R+\|\mathbf{x}\|,
\\
1 & r\ge R+\|\mathbf{x}\|.
\end{matrix}\right.
\end{align}

\textbf{Case 5:} If $\frac{R+r_0}{2} \le \|\mathbf{x}\| <R$, then
\begin{align}
{\mathbb{P}}\left( {d \leq r} \right) = \left\{\begin{matrix}
\frac{ {r^3}}{{ R^3-r_0^3}}  &\ 0\leq r<R-\|\mathbf{x}\|,\\
\frac{{\cal V}_{\|\mathbf{x}\|}(r,R)}{{ \frac{4}{3}\pi (R^3-r_0^3)}} &\  R-\|\mathbf{x}\| \leq r<\|\mathbf{x}\|-r_0,\\
\frac{{\cal V}_{\|\mathbf{x}\|}(r,R)-{\cal V}_{\|\mathbf{x}\|}(r,r_0)}{{ \frac{4}{3}\pi (R^3-r_0^3)}} &\  \|\mathbf{x}\|-r_0 \leq r<\|\mathbf{x}\|+r_0,\\
\frac{{\cal V}_{\|\mathbf{x}\|}(r,R)-\frac{4}{3}\pi r_0^3}{{ \frac{4}{3}\pi (R^3-r_0^3)}} &\  \|\mathbf{x}\|+r_0 \leq r<R+\|\mathbf{x}\|,\
\\
1 & r\ge R+\|\mathbf{x}\|.
\end{matrix}\right.
\end{align}

\textbf{Case 6:} If $\|\mathbf{x}\| \ge R$, then
\begin{align}
{\mathbb{P}}\left( {d \leq r} \right) = \left\{\begin{matrix}
0  &\ 0\leq r<\|\mathbf{x}\|-R,\\
\frac{{\cal V}_{\|\mathbf{x}\|}(r,R)}{\frac{4}{3}\pi{( R^3-r_0^3)}} &\  \|\mathbf{x}\| -R\leq r<\|\mathbf{x}\|-r_0,\\
\frac{{\cal V}_{\|\mathbf{x}\|}(r,R)-{\cal V}_{\|\mathbf{x}\|}(r,r_0)}{{ \frac{4}{3}\pi (R^3-r_0^3)}} &\ \|\mathbf{x}\|-r_0 \leq r<\|\mathbf{x}\|+r_0,\\
\frac{{\cal V}_{\|\mathbf{x}\|}(r,R)-\frac{4}{3}\pi r_0^3}{{ \frac{4}{3}\pi (R^3-r_0^3)}} &\ \|\mathbf{x}\|+r_0 \leq r<R+\|\mathbf{x}\|,
\\
1 & r\ge R+\|\mathbf{x}\|.
\end{matrix}\right.
\end{align}

After taking derivative of ${\mathbb{P}}\left( {d \leq r} \right)$, the proof is complete.

\end{document}